\documentclass[journal]{IEEEtran}

\usepackage{blindtext}
\usepackage{subcaption}
\usepackage{graphicx}
\usepackage{algorithm, amsmath}
\usepackage[noend]{algpseudocode}
\usepackage{stfloats}
\usepackage{booktabs}
\usepackage{hyperref}

\ifCLASSINFOpdf
\else
\fi
\usepackage{subcaption}
\usepackage{graphicx}
\usepackage{xcolor}
\newif\ifcolormarker
\colormarkertrue

\begin{document}
%
% paper title
% Titles are generally capitalized except for words such as a, an, and, as,
% at, but, by, for, in, nor, of, on, or, the, to and up, which are usually
% not capitalized unless they are the first or last word of the title.
% Linebreaks \\ can be used within to get better formatting as desired.
% Do not put math or special symbols in the title.
\title{ENERO: Efficient Real-Time WAN Routing Optimization with Deep Reinforcement Learning}

% author names and affiliations
% use a multiple column layout for up to three different
% affiliations

\author{\IEEEauthorblockN{Paul~Almasan\IEEEauthorrefmark{1},
Shihan~Xiao\IEEEauthorrefmark{2},
Xiangle~Cheng\IEEEauthorrefmark{2},
Xiang~Shi\IEEEauthorrefmark{2},
Pere~Barlet-Ros\IEEEauthorrefmark{1}, 
Albert~Cabellos-Aparicio\IEEEauthorrefmark{1}}
\IEEEauthorblockA{\IEEEauthorrefmark{1}Barcelona Neural Networking Center, 
Universitat Politècnica de Catalunya, Spain\\
}
\IEEEauthorblockA{\IEEEauthorrefmark{2}Network Technology Lab., Huawei Technologies Co.,Ltd.}}

\maketitle

\begin{abstract}

Wide Area Networks (WAN) are a key infrastructure in today's society. During the last years, WANs have seen a considerable increase in network's traffic and network applications, imposing new requirements on existing network technologies (e.g., low latency and high throughput). Consequently, Internet Service Providers (ISP) are under pressure to ensure the customer's Quality of Service and fulfill Service Level Agreements. Network operators leverage Traffic Engineering (TE) techniques to efficiently manage the network's resources. However, WAN's traffic can drastically change during time and the connectivity can be affected due to external factors (e.g., link failures). Therefore, TE solutions must be able to adapt to dynamic scenarios in real-time.

In this paper we propose Enero, an efficient real-time TE solution based on a two-stage optimization process. In the first one, Enero leverages Deep Reinforcement Learning (DRL) to optimize the routing configuration by generating a long-term TE strategy. To  enable efficient operation over dynamic network scenarios (e.g., when link failures occur), we integrated a Graph Neural Network into the DRL agent. In the second stage, Enero uses a Local Search algorithm to improve DRL's solution without adding computational overhead to the optimization process. The experimental results indicate that Enero is able to operate in real-world dynamic network topologies in 4.5 seconds on average for topologies up to 100 links.

\end{abstract}

\IEEEpeerreviewmaketitle

\section{Introduction}

In the last years, Wide Area Networks (WAN) have seen a considerable growth in network traffic and applications (e.g, live video streaming, Internet of Things), imposing new requirements on existing network technologies \cite{8610949,6964248, ellis2016communication}. Some examples of such requirements are to support sudden changes in the network traffic, to enable the deployment of applications with different requirements (e.g., low latency and high throughput) or to adapt to network topology changes (e.g., link failures, spikes of traffic). This puts pressure on the Internet Service Providers (ISP) to ensure the customer's Quality of Service and fulfill the Service Level Agreements. Therefore, ISPs are challenged to efficiently and effectively manage their WAN infrastructure to guarantee previously agreed losses, delay and throughput thresholds for different time-sensitive applications.

In order to efficiently manage WAN infrastructures operators take advantage of Traffic Engineering (TE) techniques \cite{awduche2002overview}. TE aims to efficiently manage the network resources by steering traffic to achieve a certain goal, for instance minimizing the utilization of the most congested link. In our work, the TE problem is defined by the network infrastructure, the traffic matrix, the routing and the link capacity (Section~\ref{subsec:problemstatement}). TE is a well-established topic with a large set of proposals. Initial attempts aimed at optimizing the link weights using distributed mechanisms using either heuristics or classical optimization techniques \cite{832225, xiao2000traffic, fortz2002optimizing}.

WANs have recently been \emph{softwarized}, this is referred to as SD-WAN~\cite{yang2019software}. SD-WANs offer programmability and the SDN controller has a full view over the network resources, enabling a new breed of centralized TE algorithms. A notable example is DEFO~\cite{10.1145/2829988.2787495}, which uses a centralized constraint programming algorithm to produce TE solutions in a few minutes. The centralization and softwarization of the network has allowed to achieve unprecedented TE performance~\cite{akyildiz2014roadmap}.

In this paper, we explore the feasibility of designing a Deep Reinforcement Learning (DRL) based method for solving TE problems. We propose \textit{Enero}\footnote{EfficieNt rEal-time Routing Optimization}, a real-time high performance optimization engine for solving TE problems (Section \ref{sec:design}). In addition to DRL, we use a Local Search (LS) algorithm to improve DRL's solution (Section \ref{subsec:twostagesoptim}). Intuitively, LS explores the solution space by applying small changes to the DRL's solution. In contrast to other existing solutions, our method does not require the network operators to design hand-crafted heuristics nor to use expert knowledge. 

Several works analyzed WAN's traffic behavior to study and model its behavior~\cite{lucas1997statistical, thompson1997wide,wang2021examination}. These studies found that the significant changes in traffic patterns happen frequently, on the scale of several minutes. Thus, to be able to solve a TE problem before the traffic changes significantly, we considered real-time to be in a sub-minute time scale.

One of the problems of using DRL in real-world scenarios is that it does not offer performance bounds. This means that once a DRL agent is trained, there is no way to give a minimum certainty over the DRL agent's performance. This performance bound would let network operators know when the DRL agent's performance is poor before deployment and avoid compromising the real network's behavior. Consequently, network operators typically do not feel confident to deploy such technology in a real-world network. In our work, we designed a method to offer a minimum performance certainty or bound in the DRL agent's performance(Section \ref{subsec:perfbound}). 

Another important characteristic of our solution is that it is able to adapt to changes. WAN scenarios suffer from changes constantly, physical links can be broken due to external factors and network users have different pattern behaviors that cause difficult-to-predict spikes in network's resources utilization. When such an event occurs, state-of-the-art TE solutions based on heuristics or classical optimization techniques need to start the optimization process from scratch. 

In our work, Enero is designed with a DRL agent that incorporates a Graph Neural Network (GNN)~\cite{scarselli2008graph}. GNNs are a novel family of Deep Learning techniques tailored for learning relational information. By using a GNN in the DRL agent, we enable Enero to operate efficiently over different network scenarios when the traffic matrix or the network topology changes during time.

In summary, our work makes the following contributions: 

\begin{itemize}
    \item We propose Enero, a two-staged method that leverages DRL and a Local Search (LS) algorithm to reach high quality TE solutions in real-time operation (Section~\ref{subsec:twostagesoptim}).
    \item We propose a method to offer a performance certainty or lower bound in the DRL agent's operation (Section~\ref{subsec:perfbound}).
    \item We design a DRL agent that is able to operate efficiently while link failures occur and is able to adapt to dynamic traffic matrices (Section~\ref{subsec:drlagent}).
\end{itemize}

\section{Background}

\subsection{Problem Statement}
\label{subsec:problemstatement}

\begin{figure}[!t]
  \centering
  \includegraphics[width=0.9\columnwidth]{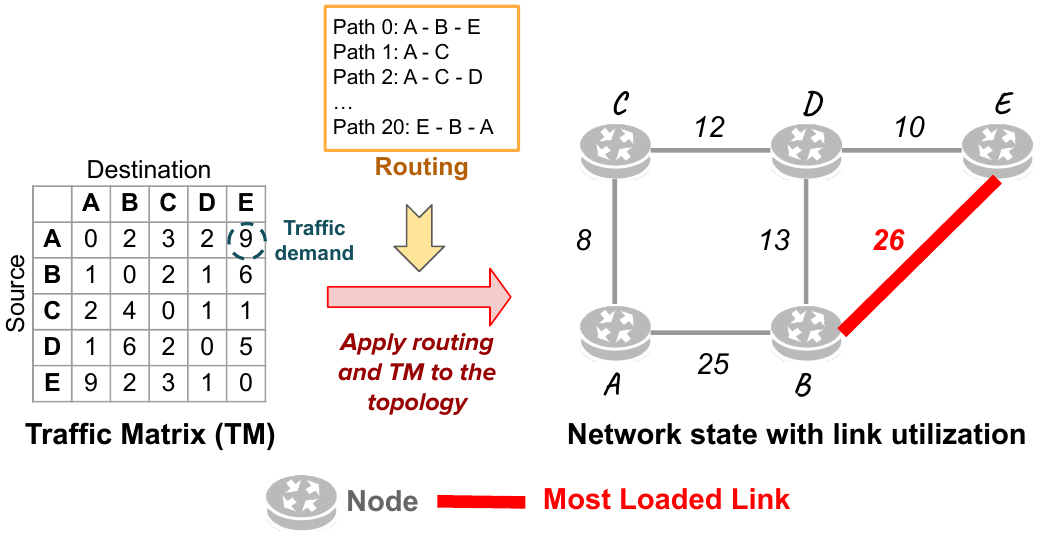}
  \caption{Problem statement overview. The routing configuration is applied to the traffic matrix and is combined with the network topology, resulting in a network state with link utilization values. Our goal is to minimize the utilization of the most loaded link.}
  \label{fig:tm_applied2Network}
\end{figure}

The problem we want to solve corresponds to the classic TE problem of minimizing the maximum link utilization \cite{832225, fortz2002optimizing, 10.1145/2829988.2787495, bhatia2015optimized}. This is because we are interested in avoiding sending packets over congested links. A congested link is where the amount of traffic crossing the link is larger than the link capacity. When this happens, the excess packets are dropped, causing packet losses. Thus, we want to minimize the most congested link and to efficiently use the network's resources. 

The TE problem is defined by a directed graph, a traffic matrix (TM) and an initial routing configuration. We abstract the real-world network topology as a directed graph, where the physical routers are represented by nodes with no features associated. Between two nodes there are always two links which correspond to the upstream and downstream links. In reality there can be multiple links between two nodes. However, we abstract from such technicality and we aggregate all the capacities into a single link for each direction. The traffic matrix indicates the volume of traffic that is being sent through the network. Specifically, the TM has size \textit{NxN} where N is the number of nodes. Each pair of nodes \textit{(s, d) with s$\in$N and d$\in$N} corresponds to a \textit{traffic demand} which is an aggregate of flows. In our optimization scenario, we do not take into account the traffic demands with \textit{s==d} (i.e., the nodes do not send traffic to themselves). Figure~\ref{fig:tm_applied2Network} illustrates how the routing and the TM are combined to obtain a network state with the link utilization values.

Initially, each traffic demand is allocated using the OSPF protocol with unitary link weights. These weights are initially assigned by the network operator using different methods (e.g., unitary weight or inverse of the link capacity). Then, the goal is to change the routing policy such that the maximum link utilization is minimized. Ideally, the final solution should decrease the link utilization in a way that the amount of traffic volume crossing the most loaded link is below the link's capacity.

We leverage Segment Routing (SR) \cite{7417124} to enable fast and efficient centralized network management. SR is a protocol that includes routing related-information in the IP packet headers. This means that each packet will have an SR path to reach a destination node. Then, SR Ingress routers encapsulate incoming packets to create a tunnel that traverses an SR path before reaching their respective destination. This SR path is composed of different segments, and in each of them, the endpoint node removes the outermost encapsulation label. This process is repeated until the packet reaches the SR Egress node. The packets within a segment are routed using the traditional OSPF routing protocol. In TE terms, SR can program detours in forwarding paths so that network packets avoid crossing congested links. Previous work showed that SR using 2-segment paths offers enough flexibility to achieve high network performance \cite{7218434}. In our work, we adapt a similar approach and we consider only one intermediate node between SR Ingress and Egress nodes. 

\begin{figure}[!t]
  \centering
  \includegraphics[width=0.99\linewidth]{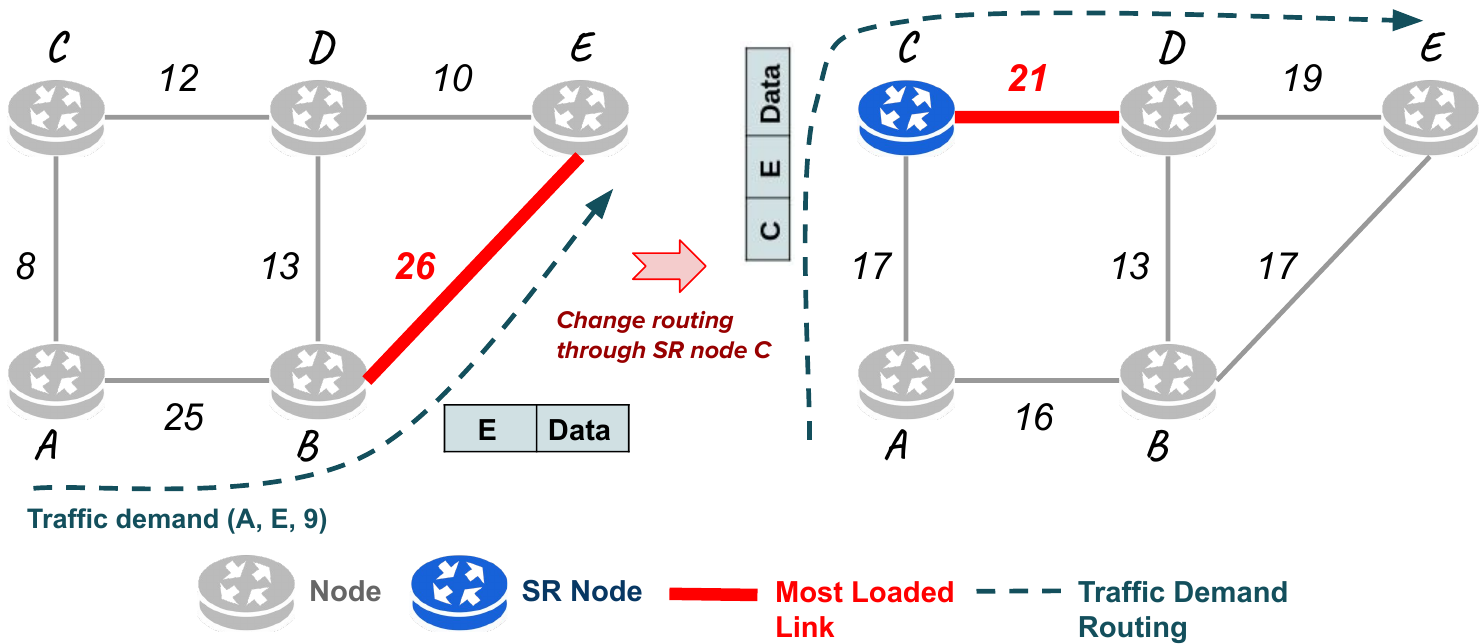}
  \caption{Single step optimization process that illustrates how changing the routing of the traffic demand (A, E) minimizes the maximum link utilization. Specifically, traffic demand (A, E) is assigned with the intermediate SR node C to program a detour and avoid link B-E.
  }
  \label{fig:ProblemStatement}
\end{figure}

Figure \ref{fig:ProblemStatement} shows an illustrative example of the TE problem we want to solve. In the figure, the overlay routing for a single traffic demand is changed. Specifically, the traffic demand that goes from node A to node E has a bandwidth of 9 and it initially uses OSPF to reach the destination. This corresponds to the left-hand side network state from the same figure. Then, a good action to minimize the maximum link utilization would be to re-route the traffic demand through the intermediate node C. This means that the SR path would be A - C - E, where C is the intermediate node. This process is repeated for all the traffic demands (i.e., all pairs of source-destination nodes), where their routing policies are changed such that the maximum link utilization is minimized.

\subsection{Shortcomings of existing solutions}

The TE problem can be formulated as an Integer Linear Programming (ILP) problem and can be solved using state-of-the-art optimizer engines such as Gurobi~\cite{gurobi} or CPLEX~\cite{cplex2009v12}. In our TE problem, the decision variables correspond to the traffic demands and the link capacities constrain the optimisation problem and define the solution space. There is at most one traffic demand for each pair of nodes. When the problem size grows (i.e., the number of nodes and links grows), the number of decision variables increases and the solution space becomes larger and more complex. In this context, TE in WAN results in a large combinatorial space where the number of possible routing configurations for each traffic demand explodes. Consequently, ILP solvers would take several weeks to find the exact solutions in WANs as they have in the order of hundreds of links and nodes~\cite{10.1145/2829988.2787495, knight2011internet}.

An alternative to ILP is the use of Constraint Programming (CP)~\cite{rossi2006handbook}. This method defines the combinatorial problem to solve with a set of decision variables (e.g., traffic demands, OSPF weights), a set of domains (i.e., potential values of the decision variables) and a set of constraints on the feasible solutions (e.g., maximum link utilization must be below a threshold). To define the constraints that limit the solution space and makes it tractable is non-trivial in WANs due to their size and complexity. In addition, the user indicates some time limit and the solver will return the best solution found within the specified time (e.g., DEFO~\cite{10.1145/2829988.2787495}). Therefore, when solving a TE problem using CP, network operators should estimate the solver's execution time needed to obtain a solution with the desired performance. However, WANs experience external events frequently (e.g., link failures, increase in traffic demand), altering the normal network behavior~\cite{lucas1997statistical, thompson1997wide,wang2021examination}. This method has the limitation of finding sub-optimal solutions if the specified time is not long enough. 

Finally, network operators can use heuristics or expert knowledge to design an algorithm to solve TE problems. In addition, they can leverage heuristics to reduce the problem dimensionality by pruning the solution space, and then use a traditional method to solve the smaller problem (e.g., CP, ILP). In the last years, WANs' size and traffic have been growing by almost doubling every year \cite{8610949,6964248, ellis2016communication}, raising the complexity of efficient network operation. As a result, the design of high performance heuristics for TE became more challenging for humans, and with a higher cost for network operators. In addition, human experts typically use trial-and-error processes that can take several months, which does not scale with recent trends in WANs.

\subsection{Deep Reinforcement Learning for Traffic Engineering}

DRL is a technology capable of modeling future rewards. This means that DRL can optimize the routing configuration taking into account the future. That is to say, DRL can learn a long-term routing policy by taking into account the future expected rewards. For example, to change a routing policy of a traffic demand might not lead to an immediate minimization of the maximum link utilization but to a delayed one that the DRL agent will observe later in the future. This is contrary to heuristics where they can not establish a relationship between local decisions (e.g., change a routing configuration for a traffic demand) and long-term strategies to solve an optimization problem (e.g., minimize the most loaded link), leading heuristics to achieve sub-optimal performance. The long-term planning capabilities make DRL a key technology for solving the TE problem.

The TE problem can be seen as a combinatorial problem where traffic demands are assigned to routing policies such that the utilization of the most loaded link is minimized. The difficulty of combinatorial problems can make the DRL reach sub-optimal solutions. The reason behind this is that when the DRL agent makes a bad decision, it has no way to undo it and explore other actions. To solve this issue, we incorporated a low computational overhead optimization step that is executed after the DRL's agent optimization process.

\section{Design}
\label{sec:design}

Enero is a two-stage method for real-time routing optimization that combines DRL and LS. In the first stage, Enero leverages DRL to find a good initial solution to the TE problem by taking into account future traffic demands. Recall that we consider a traffic demand as a source-destination node pair with a bandwidth that represents an aggregate of flows between the node pair. In the second stage, Enero tries to improve DRL's solution using a LS technique. In our work, the LS step implements the hill climbing heuristic that behaves in a greedy way by making incremental changes to the DRL solution. 

Intuitively, LS explores the solution space by applying small changes to the initial configuration or solution. The motivation behind the combination of DRL with LS is to leverage DRL's long-term planning capabilities and to improve DRL's solution using LS. Combining DRL with traditional optimization techniques has shown to achieve high performance in complex scenarios \cite{cappart2020combining,zhang2020cfr}. We believe that DRL and LS complement each other, increasing the performance of the resulting solutions.

The number of traffic demands grows quadratically with the number of nodes in a network. For instance, in a topology with 30 nodes there are $30*29=870$ traffic demands whose routing needs to be reconfigured to solve the TE problem. Ideally, we would like to take into account all traffic demands in our TE optimization problem to ensure that our solver can find the best routing configuration. However, the solution space becomes intractable for large TE problem instances and computationally expensive even when using heuristics. Inspired by \cite{zhang2020cfr}, we decided to take a subset of these traffic demands. These are called \textit{critical demands} and they are selected from the set of traffic demands crossing the 5 most loaded links. We initially performed some experiments where we optimized selecting different percentages of the traffic demands (i.e., 10, 15, 20 and 50). The results showed that taking 15\% of the critical demands offered the best trade-off between computation time and performance.

\subsection{Two-stage optimization}
\label{subsec:twostagesoptim}

The complexity of the combinatorial problem can make the DRL agent achieve sub-optimal routing configurations. This is because, on the contrary to some existing solutions that use backtracking (e.g., DEFO~\cite{10.1145/2829988.2787495}), the DRL agent has only one shot to converge to the optimal solution (i.e., a single iteration over all traffic matrices). To solve this issue, we improve DRL's solution using a LS technique without adding a large computational overhead. 

\begin{figure*}[!t]
  \centering
  \includegraphics[width=0.8\linewidth,height=5.7cm]{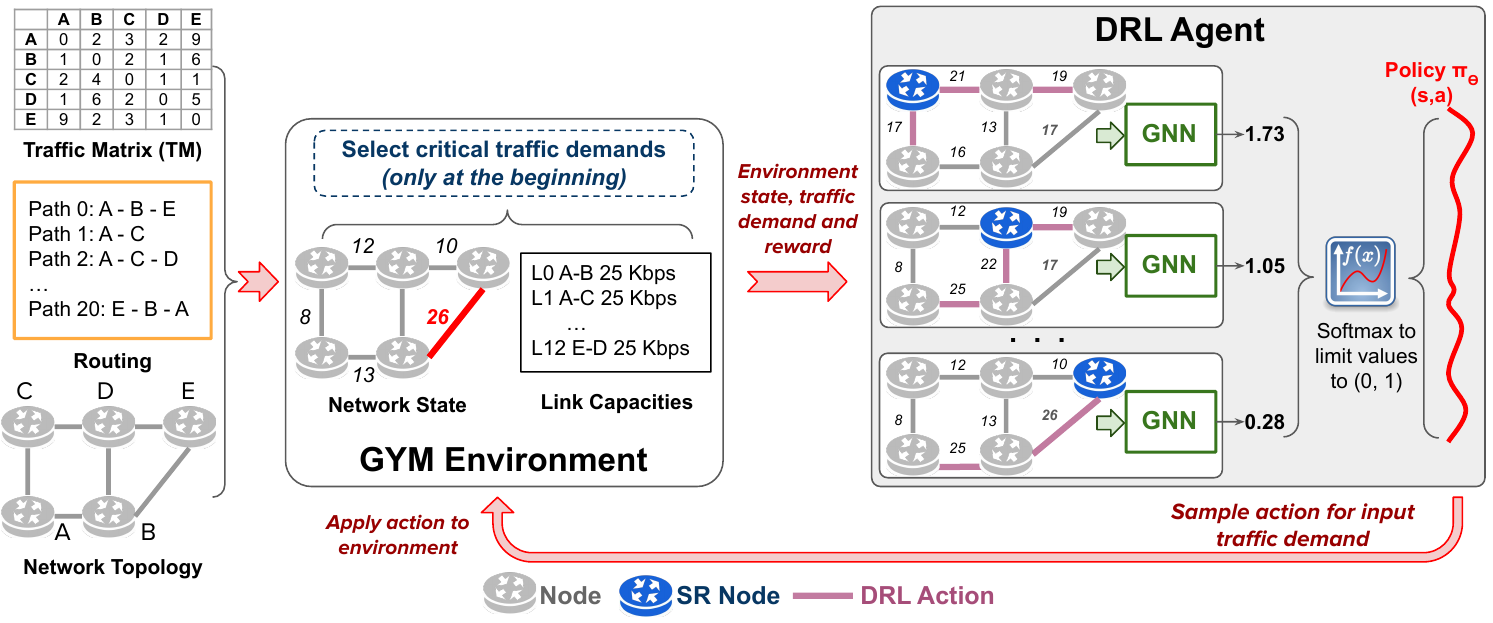}
  \caption{DRL setup overview. The \textit{critical demands} are computed in the GYM Environment at the beginning of an episode. Then, the DRL agent iterates over them and for each demand, the agent explores the action space by marking the links for each SR path. Afterwards, a GNN takes the graphs with the actions marked and outputs a probability distribution. The action to perform is sampled from the distribution and applied to the environment.}
  \label{fig:DRL_action_space}
\end{figure*}

The LS step implements a hill climbing heuristic. This method makes small incremental changes to the DRL's solution, trying to find new TE solutions that are fundamentally close to DRL's resulting configuration. Specifically, LS iterates over all traffic demands and all possible SR paths, trying to find which is the best configuration that minimizes the maximum link utilization. Similarly to the DRL case, LS iterates only over 15\% of the critical traffic demands. We decided to adapt LS in the second stage for being an anytime optimization technique. This means that the LS search process can be stopped at any moment and the result returned will always be a valid one. 

\subsection{Performance Lower Bound}
\label{subsec:perfbound}

Even though DRL is a key technology to learn long-term strategies, it can still make mistakes. DRL is a data-driven method and when evaluated in out-of-distribution data (i.e., data totally different than the one used in the learning process), it is to be expected that the performance will degrade. In our TE problem this can happen due to different bandwidth scales in the traffic matrices, due to different extreme topologies that can radically change the action space or because of the high complexity of exploring the solution space. 

To solve this problem and to enable the deployment of the DRL technology in real-world scenarios, we had to give some minimum performance certainty for the DRL agent. With this lower bound, the network operator can know for certain the DRL agent's performance before deploying it on the real network. To do this, the DRL agent starts the optimization process from a predefined routing policy. In our paper we consider OSPF as the starting routing policy. Then, it starts the optimization process and changes the routing configuration of the critical traffic demands. If the DRL agent is not capable of minimizing the maximum link utilization, it returns the initial routing configuration to the LS stage. Enero is designed to allow the starting routing policy to be initialized using any routing policy (e.g., expert-knowledge, heuristic-based routing policy).

\subsection{Deep Reinforcement Learning Agent}
\label{subsec:drlagent}

The DRL setup can be described by defining the environment state, the action representation and the reward. The \textit{environment state} includes the network topology with the links' features (i.e., link capacity and utilization). When the DRL agent performs an action (i.e., applies a new routing policy to a given traffic demand), the link's utilization is updated. The \textit{action} is represented directly on the network topology by marking the links that are part of the action. In other words, the links of the path going from a source node to a SR intermediate node and from this to the destination node are marked with a flag.
All the nodes from the topology can be SR intermediate nodes. This process is repeated for each possible action of the current traffic demand whose routing needs to be changed. The DRL's GNN is then in charge of processing these graphs (i.e., one graph per action where the action is marked directly on the links) and will output a probability distribution over the actions. Finally, the reward is the difference between the maximum link utilization between two steps. This difference is relative to the link capacities. Figure~\ref{fig:DRL_action_space} shows an overview of the DRL setup and the operation process.

\subsection{Workflow}
\label{subsec:workflow}

\begin{figure}[!t]
  \centering
  \includegraphics[width=0.95\linewidth]{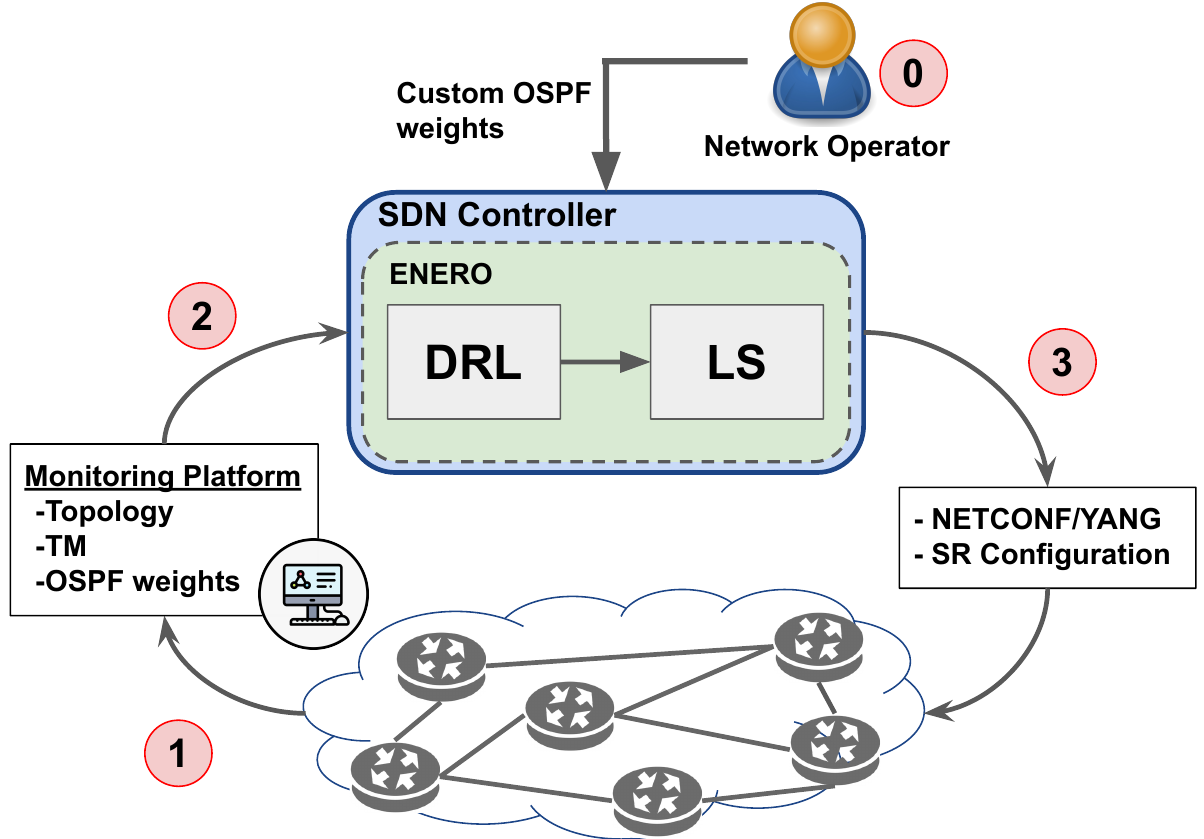}
  \caption{Enero's workflow.}
  \label{fig:EneroWorkflow}
\end{figure}

Enero is an optimization engine that is placed in the SDN controller. It takes as input the network topology, the TM and the initial routing configuration. When the SDN controller detects a change in the network (e.g., traffic matrix changed, link failure), it executes Enero to start the optimization process for the new scenario. This process finishes in under a minute, enabling real-time operation.

Figure \ref{fig:EneroWorkflow} shows \textit{Enero's} step-by-step workflow. At the beginning (Step 0), the Network Operator defines the initial OSPF weights. These weights are used to initially route the traffic demands and to route the demands within SR segments \cite{7218434}. Their values can be assigned either by the network operator using heuristics and expert knowledge or by using some well-established OSPF weights initialization (e.g., unitary weight values, inverse of the link capacity).

Once the initial routing policy is defined, a monitoring platform is in charge of retrieving the relevant information for the TE optimization problem (Step 1). This information consists of the network topology, the TM and the OSPF weights. Then, Enero takes this information (Step 2) and starts the optimization process. When the process finishes, the routing configuration (i.e., per-demand SR intermediate node assignment) is pushed down to the Data Plane (Step 3). This means that each traffic demand is going to be assigned a SR intermediate node. When there is some change in the Data Plane (e.g., the topology or the TM changed), the monitoring platform will detect these changes and will launch Enero again to optimize the new scenario. There are many efforts put on the design of fast and efficient monitoring platforms and we consider it to be outside the scope of this work \cite{10.1145/3230543.3230544,10.1145/3341302.3342076,10.1145/3098822.3098831}.

\subsection{Training Algorithm}

The DRL agent training is an iterative process that takes as input a network topology, a set of traffic matrices, the links' features and the initial OSPF weights defined by the network operator. Then, the DRL agent will learn how to optimize over the given routing configuration and for different TMs. To do this, the DRL agent iterates over the traffic demands following a decreasing bandwidth order, changing the routing policy for each demand. This means that for each traffic demand, the DRL agent will assign the best SR intermediate node before reaching the destination node. This process can be seen as changing the direct shortest path from the source node to the destination node by creating a detour. This is a trial-and-error process where at the beginning the agent will explore different routing configurations, and as the training advances, the agent will tend to exploit more of the action space. 

Algorithm \ref{alg:operation} shows the pseudo-code of the actor-critic training process. For the sake of simplicity, the pseudo-code describes the training process using a single network topology. The same process can be applied to multiple topologies by repeating the lines \ref{line:init} to \ref{line:store_res} for each topology. The training process starts in line \ref{line:for_loop} and finishes when the number of training episodes \textit{E} has been reached. At the beginning of the training episode, the DRL environment is initialized (line \ref{line:init}). This means that the topology is built and the links' utilization is updated according to the initial OSPF routing policy.

\begin{algorithm}[!t]
\caption{DRL Agent Training Process}
\begin{algorithmic}[1]
\State \textbf{Input:} Network topology (G), link capacities, TMs, Initial OSPF Weights (OSPFw)
\For {\textit{i} in \textit{${0,...,E}$}}\label{line:for_loop}
\State $env, d \gets init\_env(G, TM, OSPFw)$\label{line:init}
\While {not Done}\label{line:while-init}
\State $act\_dist \gets pred\_act\_distrib(env, d)$\label{line:pred_act}
\State $c\_val \gets pred\_critic\_value(env)$\label{line:critic_value}
\State $a \gets choose\_action(act\_dist)$\label{line:act_sampling}
\State $d, Done, r \gets step(a, d, env)$
\State $store\_results(act\_dist, c\_val, a, d, Done, r)$\label{line:store_res}
\EndWhile
\State $c\_val \gets pred\_critic\_value(env)$\label{line:critic_value2}
\State $ret, adv \gets compute\_GAE(c\_val', r')$
\State $a\_loss \gets compute\_actor\_loss(adv)$
\State $c\_loss \gets compute\_critic\_loss(ret)$
\State $total\_loss \gets a\_loss + c\_loss - entropy$ 
\State $grads \gets compute\_gradients(total\_loss)$
\State $clip\_gradients(grads)$
\State $apply\_gradients(grads)$
\EndFor
\end{algorithmic}
\label{alg:operation}
\end{algorithm}

The loop from line \ref{line:while-init} indicates the iteration of the DRL agent over the critical traffic demands. In each loop iteration, the DRL agent tries to change the routing policy of a single traffic demand (i.e., assign an SR intermediate node). In line \ref{line:pred_act} the DRL agent uses the GNN to output a probability distribution over the action space. Then, the critic network predicts the value of the current state.

The DRL agent uses a random sampling of the action distribution to pick the action to perform (line \ref{line:act_sampling}). During evaluation, the sampling is changed by taking the action with higher probability. Then, the selected action is sent to the environment to be applied over the current network state and to update the link's utilization. In line \ref{line:store_res}, the agent stores all the intermediate results that will later be used to compute the losses.  

The next step is to compute the Generalized Advantage Estimates (GAE), which is a method to reduce variance in policy gradient algorithms \cite{schulman2015high}. Then, the actor and the critic losses are computed~\cite{schulman2017proximal}. These are then combined in a sum and subtracted with the entropy term, used to guide the exploration during training \cite{pmlr-v97-ahmed19a}. Finally, the gradients are computed and clipped to avoid the policy to change too much for a given training step, and they are applied to the actor and critic networks.

\section{Experimental Results}

In this section, we first describe the implementation details and the methodology used to obtain the datasets and to train the DRL agent. Then, we made an experimental study to see the performance gap between DRL, LS and Enero. Finally, we perform a series of experiments on different real-world network scenarios. Specifically, we want to answer the following questions:
\begin{itemize}
    \item What is the performance gap between DRL, LS and Enero for solving TE problems? (Section~\ref{sec:drl_ls_hybrid_method})
    \item How does Enero perform when the traffic matrix changes during time? (Section~\ref{sec:dynamic_tm})
    \item What is Enero's performance when the topology changes as a result of link failures? (Section~\ref{sec:link_failures})
    \item What is Enero's performance and execution cost compared with state-of-the-art TE solutions? (Section~\ref{sec:operation_perf_cost})
\end{itemize}

All the experiments were executed on off-the-shelf hardware without any specific hardware accelerator or high performance software optimization engine. Specifically, we used a machine with Ubuntu 20.04.1 LTS with processor AMD Ryzen 9 3950X 16-Core Processor.

\subsection{Implementation}

Enero is implemented in Python, with the exception of the DRL part (training and evaluation) that was implemented using Tensorflow \cite{tensorflow2015-whitepaper} and the DRL environment that was implemented using the OpenAI Gym framework~\cite{brockman2016openai}. The DRL agent is trained using the PPO algorithm~\cite{schulman2017proximal}. The LS stage is implemented totally in Python except for some operations where it uses the Numpy library \cite{harris2020array}. The LS execution cost could be improved by using more efficient libraries (e.g., Cython~\cite{10.1109/MCSE.2010.118}), which were left as future work. For some graph-related operations the NetworkX library~\cite{SciPyProceedings_11} is used. Table \ref{table:hyperparametertuning} shows the hyperparameters used during Enero's DRL agent training stage. Enero's code is publicly available\footnote{\href{https://github.com/BNN-UPC/ENERO}{https://github.com/BNN-UPC/ENERO}}.

\begin{table}[!t]
\centering
\begin{tabular}{lll}
\toprule
Hyperparameter & Value \\
\toprule
 GNN Hidden State & 20 \\
 Message Passing Steps & 5 \\
 Evaluation Episodes per Topology &  20 \\
 Training Epochs &  8 \\
 \% critical demands &  15\% \\
 Mini-batch size &  55 \\
 Learning Rate &  0.0002 \\
 Decay Rate (Decay Steps) &  0.96 (60) \\
 Entropy Beta (After 60 Episodes)  &  0.01 (0.001) \\
 GAE Gamma, Lambda &  0.99, 0.95 \\
 Gradient Clipping Value &  0.5 \\
 Actor L2 Regularization &  0.0001 \\
 Readout Units &  20 \\
 Activation Function &  Selu \\
\bottomrule
\end{tabular}
\caption{Enero hyperparameter configuration.}\label{table:hyperparametertuning}
\end{table}
\subsection{Methodology}

\subsubsection{Traffic Matrices}
The traffic matrices were generated using a uniform distribution. This means that the bandwidth values from the traffic demands were uniformly distributed from 0.5 to 1. Then, we scaled this value to obtain the TM's bandwidths in Kbps and to have the same unit for both bandwidth and link capacities. Each network topology had a total of 150 TMs. 

\subsubsection{Network Topologies}
We obtained the network topologies from the TopologyZoo dataset \cite{knight2011internet}, which contains real-world topologies from network operators. Specifically, we took all topologies up to 100 links and 30 nodes, resulting in a total of 74 topologies. From these topologies, only 3 of them were used in the DRL agent's training process. In our TE problem we only consider the link capacities, which means that nodes do not have any features associated. 

\subsubsection{DRL Agent Training}

In all the experiments from this paper we are always evaluating the same DRL agent. This means that we have only trained \textit{a single DRL agent} and incorporated it into Enero. To train the DRL agent, we arbitrarily picked 3 network topologies (i.e.,  BtAsiaPac, Garr199905 and Goodnet topologies) from the 74 topologies extracted from the Topology Zoo dataset. We split the original 150 TMs from each topology into 100 TMs for training and 50 TMs for evaluation. During training, the DRL's agent performance evolution is evaluated on the BtAsiaPac, Garr199905 and Goodnet topologies after every training step. Specifically, the agent is evaluated on 20 TMs uniformly sampled from the evaluation split for each topology.

\subsubsection{Comparison Baselines}

We compare Enero with three baselines which together represent widely used heuristics and close-to-optimal solutions. The first baseline is the Shortest Available Path (SAP) heuristic. SAP starts with the empty network and iterates over all traffic demands in decreasing order of bandwidth. This is done to allocate the bigger and critical traffic demands first. Then, each traffic demand is routed using the path with the highest available bandwidth. The second baseline corresponds to a LS algorithm. Specifically, we implemented the hill climbing search to improve an initial routing configuration in a greedy fashion. Similarly to Enero, this method starts in the same routing configuration using the OSPF protocol and tries to minimize the maximum link utilization. This is an iterative process where in each step applies the routing policy of the traffic demand that minimizes the maximum link utilization. This process finishes when the maximum link utilization does not improve anymore. 

\begin{figure}[!t]
  \centering
  \includegraphics[width=0.8\linewidth, height=4.4cm]{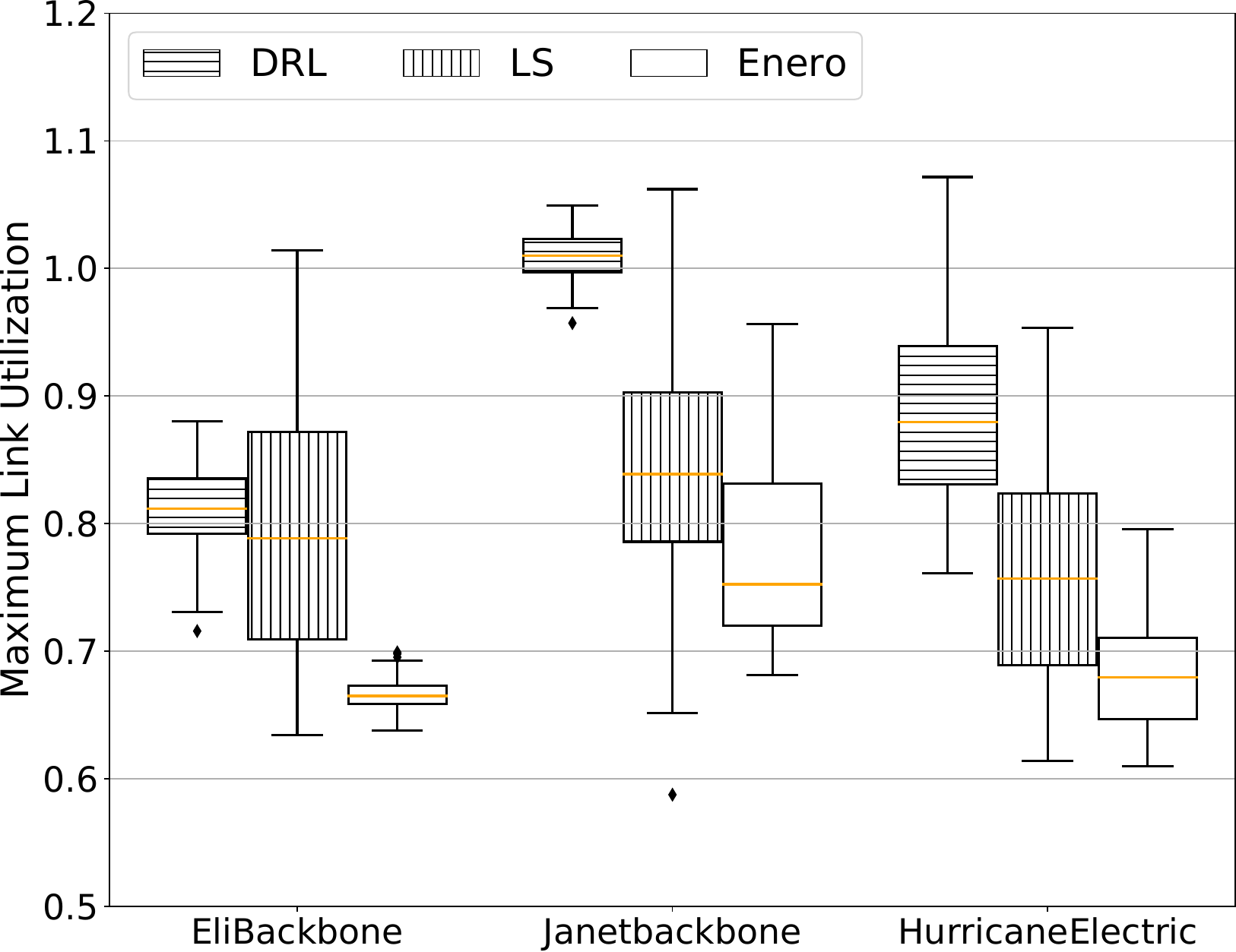}
  \caption{Performance of LS, DRL and Enero for the EliBackbone, Janetbackbone and HurricaneElectric topologies. 
  }
  \label{fig:drl_ls_enero_perf}
\end{figure}

To compute the optimal solution for our TE problem it would require weeks of computation using ILP. As it is not feasible to do that, we chose DEFO \cite{10.1145/2829988.2787495} as our close-to-optimal baseline. In particular, we took the implementation from \cite{gay2017repetita} and adapted it to have at most one intermediate SR node per traffic demand. DEFO is a CP-based solution and if left enough time executing it provides a close-to-optimal solution. This is the reason why we left DEFO executing for 180 seconds in all of our experiments. Following the recommendations from the experiments in the original paper  \cite{10.1145/2829988.2787495} on very large topologies (i.e., a few hundreds of links and more than 6.000 traffic demands to optimize), we expect that 180 seconds is enough to find close-to-optimal solutions in our topologies (i.e., we have topologies of up to 100 nodes and 900 traffic demands). 

DEFO uses Equal-cost multi-path routing (ECMP) to route the traffic demands. This enables DEFO to divide the traffic demands among multiple paths, achieving a better traffic distribution and a lower link utilization. In our problem setup the traffic demands are routed using solely a single path, creating a natural gap between DEFO and Enero's performance. We left the task of enabling Enero to optimize using ECMP for future work.

\subsection{DRL and LS Hybrid Method}
\label{sec:drl_ls_hybrid_method}

\begin{figure}[!t]
  \centering
  \includegraphics[width=0.94\linewidth]{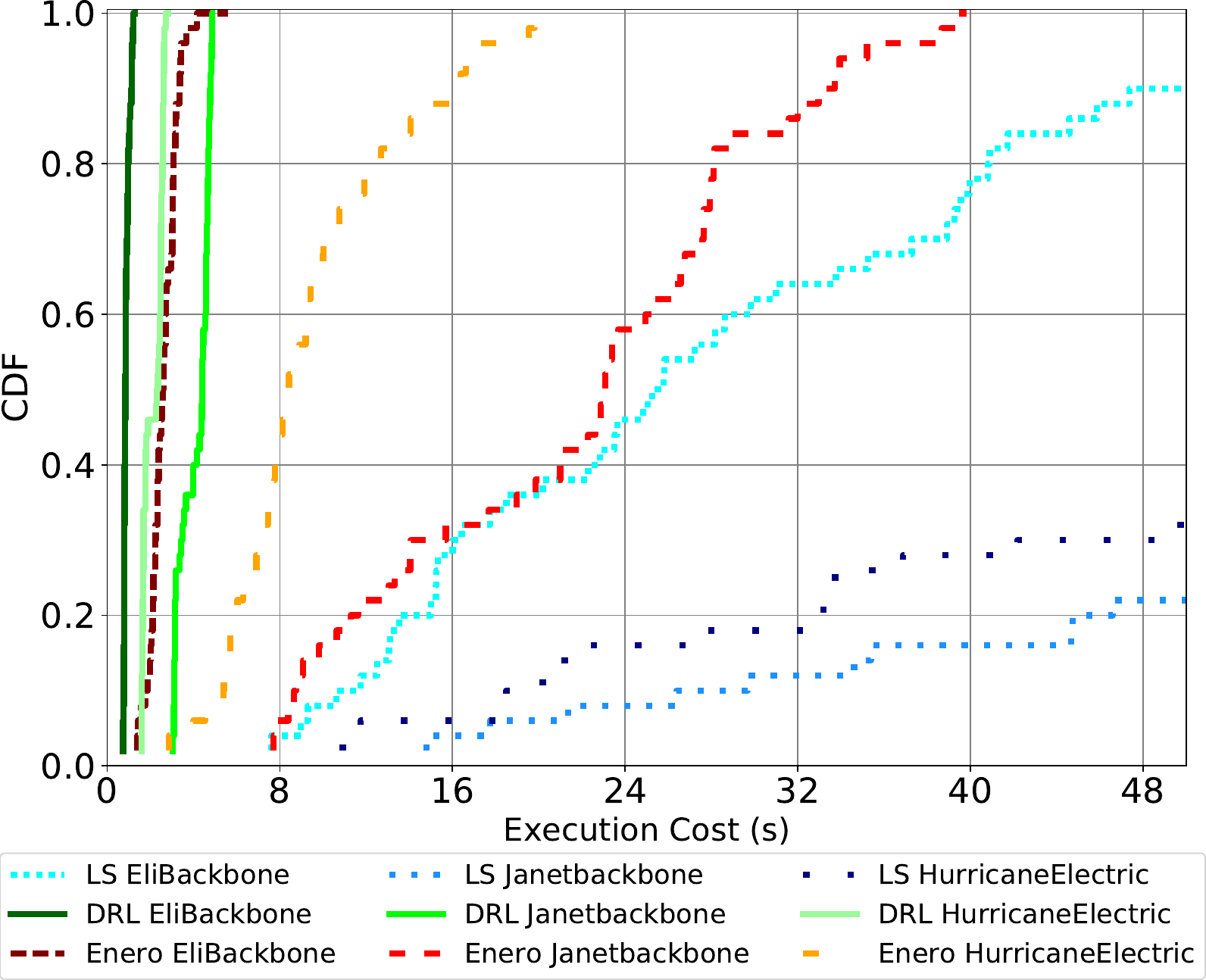}
  \caption{Execution cost of LS, DRL and Enero for the EliBackbone, Janetbackbone and HurricaneElectric topologies. Best viewed in color.}
  \label{fig:drl_ls_enero_cost}
\end{figure}

In this section we want to demonstrate the capabilities of combining DRL with LS. To do this we studied the performance and execution cost of DRL and LS individually and compared them with Enero. In the experiments, we evaluated the DRL agent, the LS algorithm and Enero on three network topologies using 50 TMs per topology. Figures \ref{fig:drl_ls_enero_perf} and \ref{fig:drl_ls_enero_cost} show the resulting performance and the CDF of the execution cost respectively. Notice that the topologies from these figures were not seen by the DRL agent during the training process. 

The experimental results indicate that DRL has a reasonably good performance  in all three topologies. This is because it can minimize the maximum link utilization from $\approx$1.1 to below 1 for EliBackbone and HurricaneElectric topologies and to $\approx$1 for the Janetbackbone topology. LS can minimize the maximum link utilization in all three topologies, obtaining better performance than DRL. However, the CDF from Figure \ref{fig:drl_ls_enero_cost} indicates that the DRL is extremely fast while LS takes a considerable amount of time (up to minutes).

To demonstrate the capabilities of combining DRL with LS we also plot in Figure~\ref{fig:drl_ls_enero_perf} and \ref{fig:drl_ls_enero_cost} Enero's performance and execution cost. The results indicate that Enero reaches better TE solutions than DRL and LS in all three topologies while the execution time is below 40 seconds for the Janetbackbone topology. Notice that the Janetbackbone topology is a large topology with 812 traffic demands whose routing policy needs to be optimized, which explains the larger execution times.

\subsection{Dynamic Traffic Matrix}
\label{sec:dynamic_tm}

\begin{figure*}[!t]
 \centering
 \subfloat[EliBackbone]{%
   \includegraphics[width=0.3\linewidth]{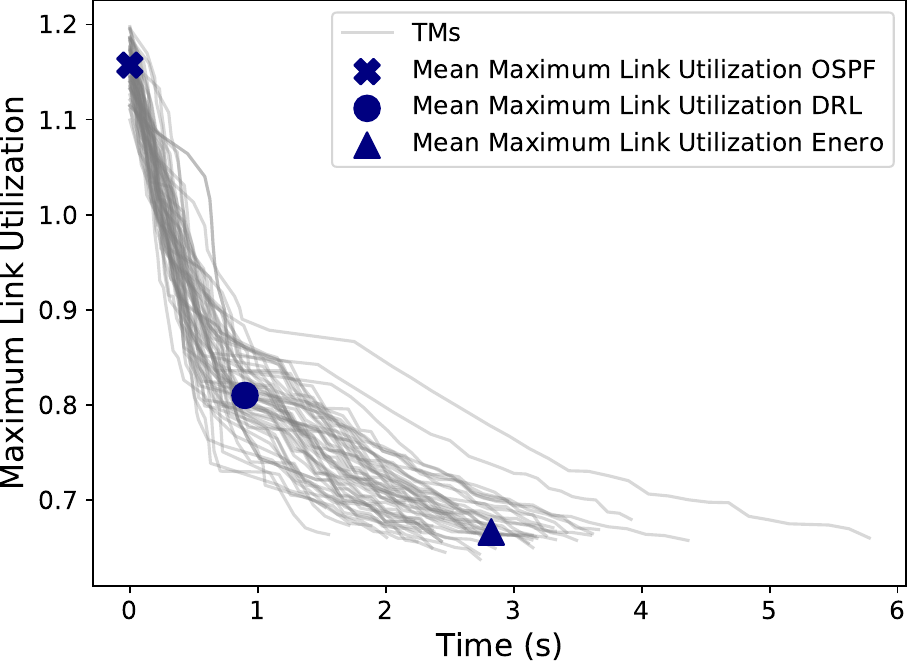}
 }
 \hfill
 \subfloat[Janetbackbone]{%
   \includegraphics[width=0.3\linewidth]{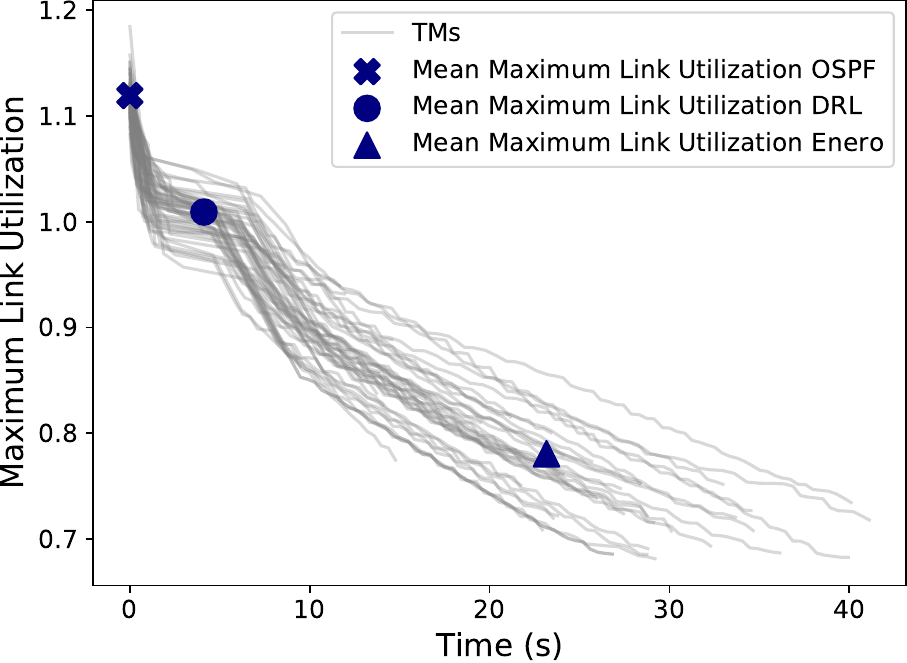}
 }
 \hfill
 \subfloat[HurricaneElectric]{%
   \includegraphics[width=0.3\linewidth]{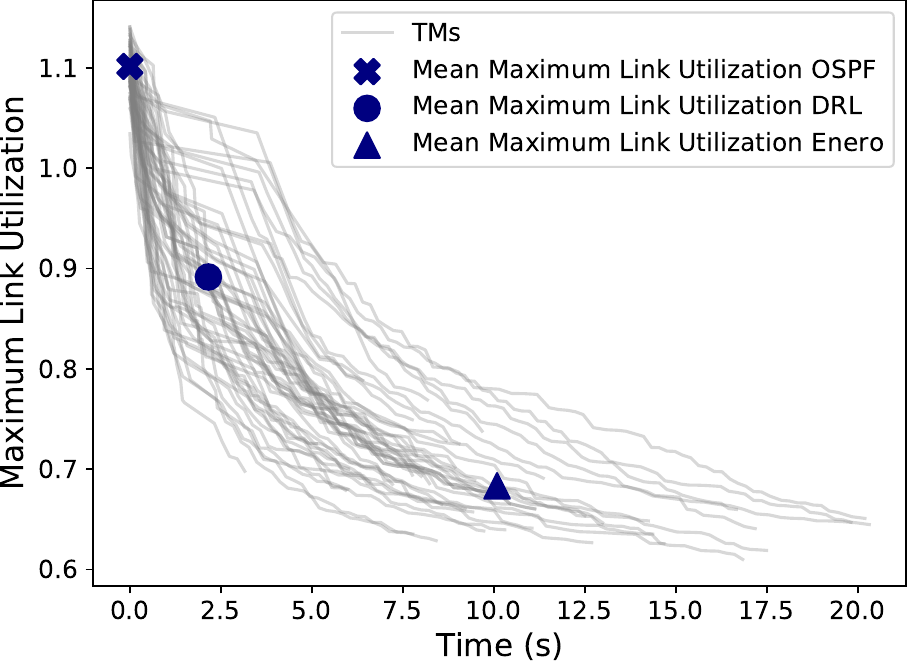}
 }
 \caption{Dynamic traffic matrix scenario. Enero evaluation on different real-world network topologies. For each topology, we evaluated over 50 different TMs. Notice that the topologies from this figure were not seen by the DRL agent during the training process.}
 \label{fig:dynamic_traffic}
\end{figure*}

In this scenario we want to evaluate Enero's performance when the traffic matrix changes during time. In our experiments we took the extreme case where every 60 seconds the entire TM changes. The reason behind this is to simulate the worst-case scenario where Enero must re-compute the solution to the TE problem from scratch. We repeat this process until the TM has changed 50 times.

Figure \ref{fig:dynamic_traffic} shows the evaluation results on three network topologies with 50 TMs per topology. Each line indicates the progress of the maximum link utilization while Enero is solving the TE problem for a given TM. In reality, the lines should be concatenated one after another but for visualization purposes we aggregated all the events where the TM changed into a single figure per-topology. From the same Figures we can observe Enero's two-stage optimization process. When the monitoring platform detects a change in the TM (see Section \ref{subsec:workflow}), Enero uses the pre-defined OSPF routing policy and then starts the optimization process. We can appreciate that in all topologies the DRL agent quickly finds a good TE solution and then LS improves it. Notice that the topologies are different from those used during the DRL agent training process. This showcases Enero's capabilities to perform TE on different network topologies (than those seen during training) and with dynamic changes in the TM.

\begin{figure*}[t]
 \centering
 \subfloat[EliBackbone]{%
   \includegraphics[width=0.3\linewidth]{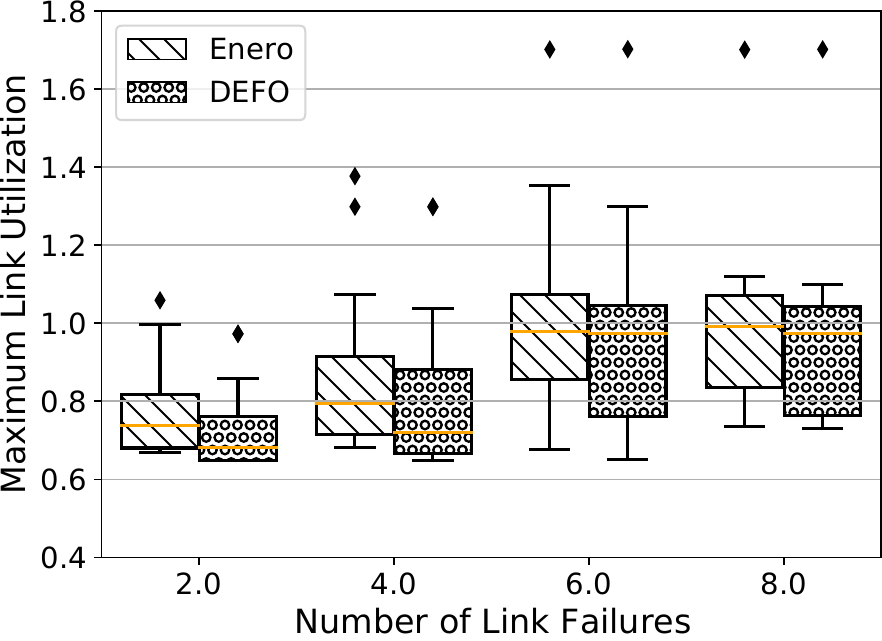}
 }
 \hfill
 \subfloat[Janetbackbone]{%
   \includegraphics[width=0.3\linewidth]{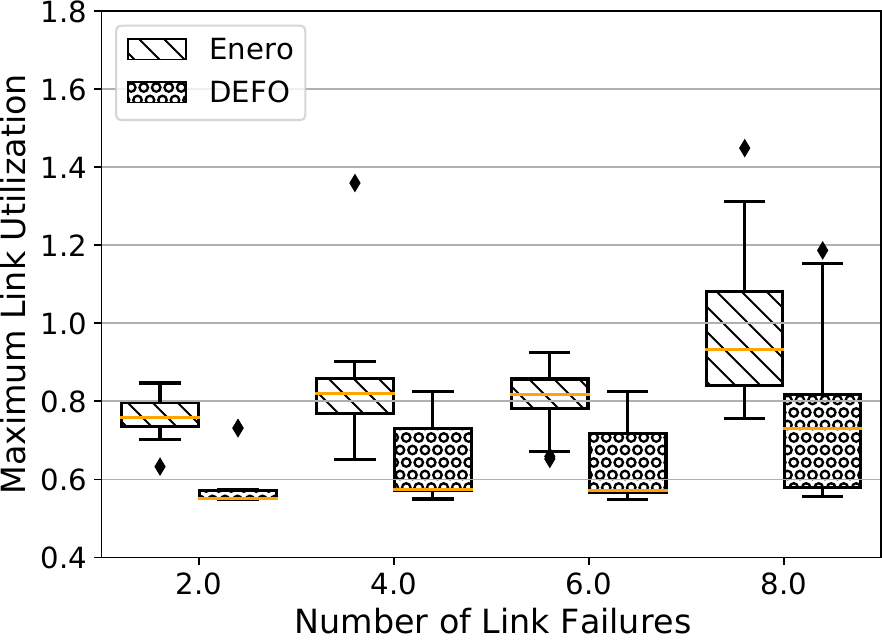}
 }
 \hfill
 \subfloat[HurricaneElectric]{%
   \includegraphics[width=0.3\linewidth]{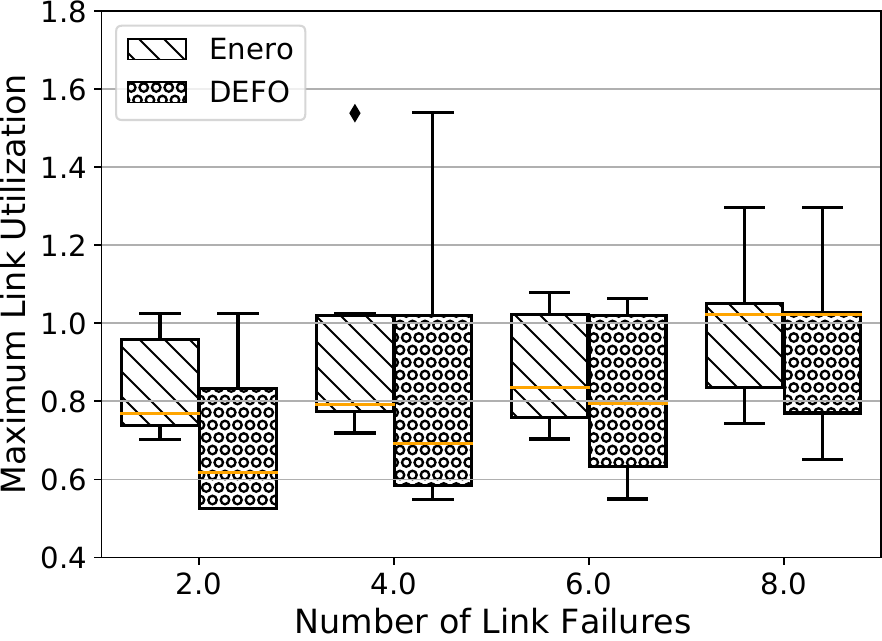}
 }
 \caption{Link failures scenario. For each number of link failures there are 20 different topologies and we evaluated using 50 TMs for each topology.}
 \label{fig:link_failure}
 \vspace{-0.3cm}
\end{figure*}

\subsection{Link Failures}
\label{sec:link_failures}

In this experiment we evaluated Enero's capabilities to react to changes in the network topology resulting from link failures. We simulated link failures by randomly removing links from the topology in each of the evaluation topologies. We made sure that there are no two topologies that are the same after removing some links. For each logical link in the topology, there are the upstream and downstream links. To ensure network connectivity, when we drop a link we drop both upstream and downstream links. We simulated up to 8 link failures in total where for each failure there are 20 different topologies and for each topology there are 50 TMs.

To make the experiments more challenging, we used the original TMs from the topologies. In other words, the bandwidths from the TMs remained the same while link failures were happening. This means that while the traffic demands did not change, link failures forced the network to have less and less resources to accommodate the original TMs.  

Figure \ref{fig:link_failure} shows Enero's results after optimization for each link failure together with the results from DEFO and SAP baselines. Because the TMs did not change and the topology had less resources to accommodate the bandwidths, the maximum link utilization should be increasing when links from the topology fail. The results indicate that Enero's performance has a similar behavior to DEFO regardless of the number of link failures. Recall that DEFO is our close-to-optimal baseline which has been executed during 180 seconds and uses ECMP to split the traffic demands among multiple paths.

\subsection{Operation Performance and Cost}
\label{sec:operation_perf_cost}

In this experiment we wanted to evaluate Enero's performance while operating on a set of real-world topologies. To do this, we took all topologies from the TopologyZoo dataset that had up to 100 links and 30 nodes. This made a total of 74 topologies, from which only 3 of them were used in the DRL agent's training process. Figure \ref{fig:zoo_tops} shows the evaluation results over all 74 topologies. Specifically, in Figure \ref{subfig:zoo_perf} we plot the relative performance with respect to the LS baseline. The topologies from 20 to 37 are ring, star or line topologies where there is no room for optimization. This explains why all the baselines have exactly the same performance. Figure \ref{subfig:zoo_cost} shows the execution cost of all the baselines. As a reminder, DEFO was set to execute for 180 seconds to ensure a close-to-optimal solution. The results indicate that Enero is capable of obtaining better performance than the SAP and LS baselines and in most of the topologies has a similar performance to DEFO. In addition, Enero's execution cost is small, with only 5 topologies with an operation cost of more than 20s. 

\section{Discussion}

\begin{figure}[!t]
    \begin{subfigure}[]{0.9\columnwidth}
	\includegraphics[width=0.89\linewidth,height=4.0cm]{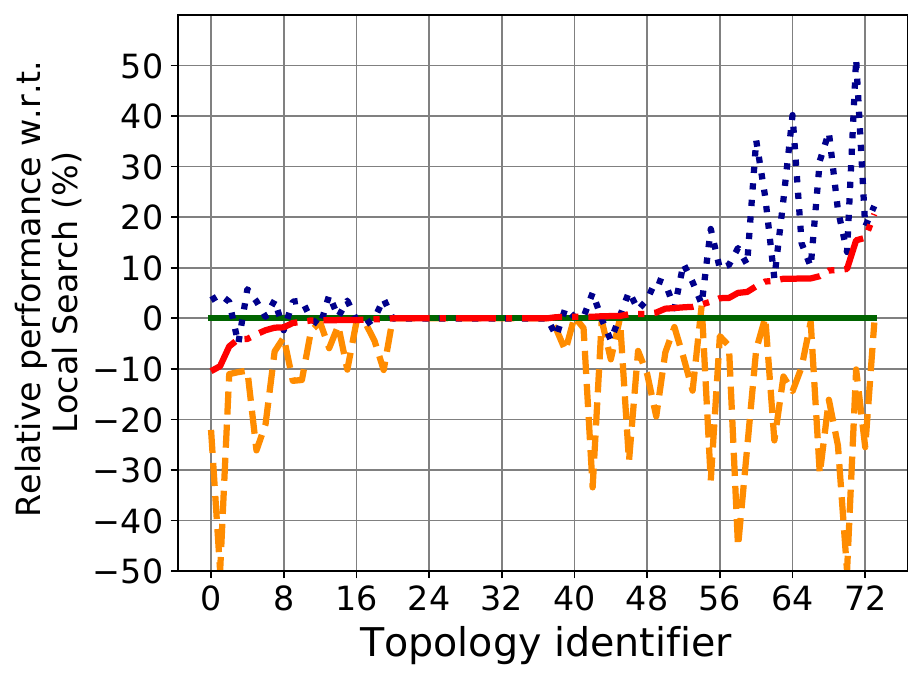}
        \caption{}
	\label{subfig:zoo_perf}
    \end{subfigure}
    \begin{subfigure}[]{0.9\columnwidth}
	\includegraphics[width=0.91\linewidth,height=4.0cm]{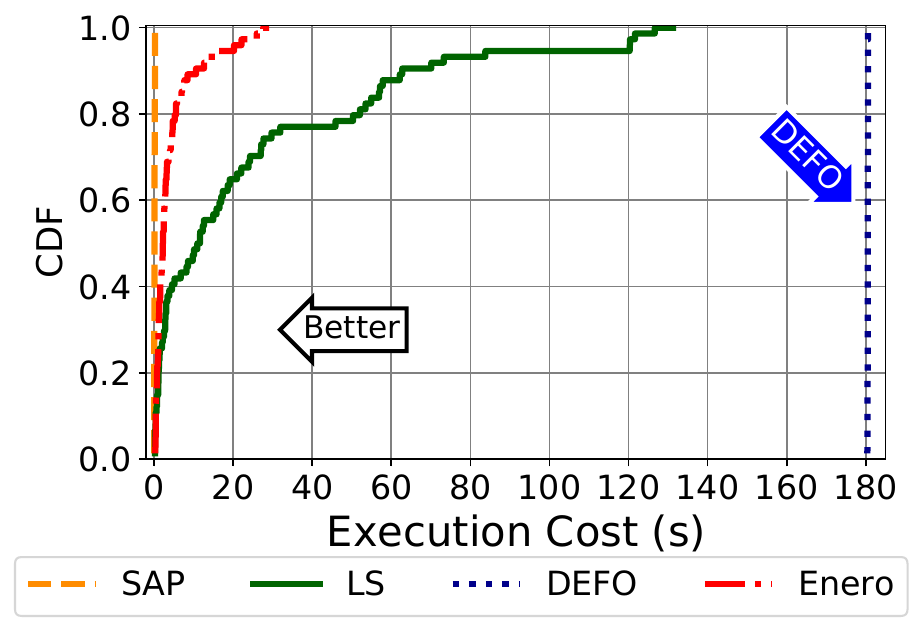}
        \caption{}
	\label{subfig:zoo_cost}
    \end{subfigure}
     \caption{Relative performance (a) and CDF of the execution cost (b) on the TopologyZoo dataset. In sub-figure (a), the topologies from 20 to 37 are ring or star topologies where there is no room for optimization.}
     \label{fig:zoo_tops}
\end{figure}

Enero is a data-driven solution that can use synthetic or real-world data to train a DRL agent to solve TE problems. This means that if we deploy our agent over topologies or TMs that are very different from those from the dataset used in the training process, we can expect our agent's performance to drop. This explains Enero's poor performance for the top left topologies from Figure \ref{subfig:zoo_perf}. Specifically, the traffic demand values are all limited by the uniform distribution between 0.5 and 1, meaning that the TMs can be discarded as the source of performance instability. Thus, we focused our attention on the network topologies and we wanted to study what is different (in connectivity terms) in the top left topologies in Figure \ref{subfig:zoo_perf}.

We identified two metrics that showcase the differences between the topologies used during training and those where Enero's performance is worse. The first one is the node degree, which indicates the number of adjacent links to a node. The second metric is the edge betweenness, which computes the portion of all pairs of shortest paths that pass through each link \textit{l} of a graph \cite{brandes2008variants}. The following equation describes the edge betweenness metric:

\begin{equation}
c(l) = \sum_{s,d \in N} \frac{\sigma(s,d | l)}{\sigma(s,d)}
\end{equation}
\label{equation:edge_betw}

where \textit{N} is the set of all nodes, $\sigma(s,d)$ is the total number of shortest paths and $\sigma(s,d|l)$ is the number of shortest paths that pass through link \textit{l}.

Table~\ref{table:zoo_metrics} shows the minimum, maximum and mean node degree and edge betweenness for each topology used during training and for the 4 topologies where Enero had worse performance. These metrics indicate that the topologies seen during training and the ones where our method performs worse are totally different. For example, the minimum and the average edge betweenness is much higher in the topologies 0, 1, 2 and 3. This indicates that the shortest paths are not well distributed and they cross the same links, making them become critical links for the TE problem. In addition, the topologies used in the training process have a higher average and a wider range of the node degree. This indicates that the nodes are more interconnected between them than in the topologies 0, 1, 2 and 3.

There are several ways to solve the out-of-distribution problem. For example, we could work with specific Deep Learning techniques such as regularization or dropout. However, the most effective way would be to add more data to the training process. This is translated to our problem by adding more topologies to the DRL's training that are different between them.  

The experimental results showed that the hybrid method of combining DRL with LS enables efficient real-time routing optimization. However, there is still room to push even further the combination of DRL with traditional optimization methods. The straight-forward approach would be to improve the LS implementation using high performance software (e.g., Cython~\cite{10.1109/MCSE.2010.118}). In addition, in our work we used a greedy approach in Enero's second stage but it could be substituted by more advanced search algorithms (e.g., CP, Genetic Algorithms). For example, the DRL's solution could be converted to constraints and then some CP solver (e.g., Gurobi~\cite{gurobi}) could find a better solution. This would ensure that the solution of the CP phase should be better than the one from the DRL agent and it would enable the second optimization stage to explore better solutions.

\begin{table}[!t]
\centering
\begin{tabular}{lll}
\toprule
Topology/Id & Node Degree & Edge Betweenness \\
\toprule
\textbf{BtAsiaPac} & (2, 24, 6.2) & (0.010, 0.067, 0.04) \\
\textbf{Goodnet} & (2, 18, 7.3) & (0.014, 0.059, 0.03) \\
\textbf{Garr199905} & (2, 18, 4.35) & (0.0435, 0.083, 0.05) \\
0 & (4, 8, 4.3) & (0.043, 0.167, 0.11) \\
1 & (2, 14, 5.23) & (0.026, 0.117, 0.07) \\
2 & (2, 8, 4.2) & (0.044, 0.164, 0.10) \\
3 & (2, 6, 4.0) & (0.067, 0.162, 0.12) \\
\bottomrule
\end{tabular}
\caption{TopologyZoo metrics. For each topology and each metric the tuple values correspond to the \textit{(min, max, mean)} values respectively. The top 3 topologies are those used during DRL's agent training process. }\label{table:zoo_metrics}
\vspace{-0.3cm}
\end{table}

\section{Related Work}

\subsection{Routing Optimization}

To find the optimal routing configuration given an estimated traffic matrix is a fundamental networking problem, which is known to be NP-hard \cite{xu2011link, 10.1145/2829988.2787495}. This problem has been largely studied in the past and we outline some of the most relevant works. The early work from \cite{832225} uses LS to find the best OSPF link weights to minimize congestion in the most congested link. In DEFO~\cite{10.1145/2829988.2787495} they propose a solution that converts high-level optimization goals, indicated by the network operator, into specific routing configurations using CP. Their problem formulation leverages SR to find the best routing configuration for each traffic demand. Within a SR path, they spread the traffic among several flows using ECMP. In \cite{gay2017expect}, the authors propose to use LS where they sacrifice space exploration to achieve lower execution times. In their design they also leverage heuristics to narrow down the LS neighborhood and to make the algorithm converge faster to good solutions. A more recent work \cite{8737424} leverages the ILP and the column generation algorithm to solve TE problems. Their solution also provides a mathematical bound to indicate how far the solution is from the optimal one. 

\subsection{Machine Learning for Communication Networks}

Recently, numerous Machine Learning-based solutions have been proposed to solve networking problems. In \cite{8485853}, they propose a generic DRL framework for TE. In their TE problem formulation, their solution consists of defining the optimal split ratios over a set of predefined paths instead of changing the paths configuration. The work from \cite{chen2020rl} applies DRL in a SDN context to solve a TE problem that maximizes the switch's throughput and minimizes the delay. The authors designed a reward that allows tuning to optimize the upward or downward throughput. In the field of optical networks, the work \cite{8761276} proposes an elaborated representation of the network state to help the DRL agent learn to efficiently route traffic demands. A more recent work \cite{almasan2019deep} proposes a different approach where the authors combine DRL and GNN to optimize the resource allocation in optical circuit-switched networks. NeuroCuts \cite{10.1145/3341302.3342221} is a DRL-based solution for solving the packet classification problem. AuTO \cite{10.1145/3230543.3230551} performs on-line traffic optimization using DRL in data center scenarios. In their work they train a DRL agent to change the queue's thresholds and another DRL agent to determine the priorities and rate limit for long flows. Decima \cite{10.1145/3341302.3342080} leverages DRL and GNN for efficient scheduling of data processing jobs in data clusters. RouteNet \cite{9109574} proposes to use GNN for network modeling and optimization.

\section{Conclusion}

Efficient real-time TE is important for network operators and ISPs to ensure network reliability when external factors can disrupt the proper network functioning. In this paper, we explored the use of DRL for solving TE problems. Specifically, we presented Enero, a method that combines DRL with LS to solve TE problems in real-time. The experimental results show that Enero is able to operate efficiently in real-world scenarios (e.g., dynamic traffic matrix, link failures). In addition, the results indicate that Enero can achieve close-to-optimal performance in less than 30 seconds for a set of arbitrary real-world network topologies. We expect our solution to inspire future work on applying DRL for solving network optimization problems. Enero's code will be publicly available.

\section*{Acknowledgment}

This work has been supported by the Spanish Government through project TRAINER-A (PID2020-118011GB-C21) with FEDER contribution, the Catalan Institution for Research and Advanced Studies (ICREA) and the  Secretariat for Universities and  Research of the Ministry of Business  and  Knowledge of the Government of Catalonia and the European Social Fund.

\bibliographystyle{IEEEtran}
\bibliography{references}

\begin{IEEEbiography}[{\includegraphics[width=1in,height=1.25in,clip,keepaspectratio]{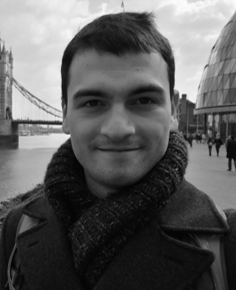}}]{Paul Almasan}
received his B.Sc. and M.Sc. in Computer Science from the Universitat Polit\`ecnica de Catalunya (UPC), Spain, in 2017 and 2019 respectively. He is currently pursuing his Ph.D. degree at the Barcelona Neural Networking Center (BNN-UPC). His research interests are focused on Graph Neural Networks and Deep Reinforcement Learning applied to network optimization.
\end{IEEEbiography}

\begin{IEEEbiography}[{\includegraphics[width=1in,height=1.25in,clip,keepaspectratio]{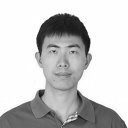}}]{Shihan Xiao} received the B.Eng. degree in electronic and information  engineering from the Beijing University of Post and Telecommunications, Beijing, China, in 2012, and the Ph.D. degree from the Department o Computer Science and Technology, Tsinghua University, China. He is currently a Senior Engineer with Huawei NetLab. His research interests include machine learning in networking, data center networking, and cloud computing.
\end{IEEEbiography}

\begin{IEEEbiography}[{\includegraphics[width=1in,height=1.25in,clip,keepaspectratio]{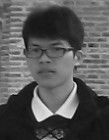}}]{Xiangle Cheng} received the M.Sc. degree in communication and information system from Southwest
Jiaotong University, Chengdu, China in 2015. He is currently a Ph.D. candidate in Computer Science at
the University of Exeter, UK. His research interests include 5G SDN/NFV, Network AI, Stochastic and Neural Combinatorial Optimization, Intelligent Wireless Networks and Mobile Computing, and Dynamic System Modelling and Performance Optimization.
\end{IEEEbiography}

\begin{IEEEbiography}[{\includegraphics[width=1in,height=1.25in,clip,keepaspectratio]{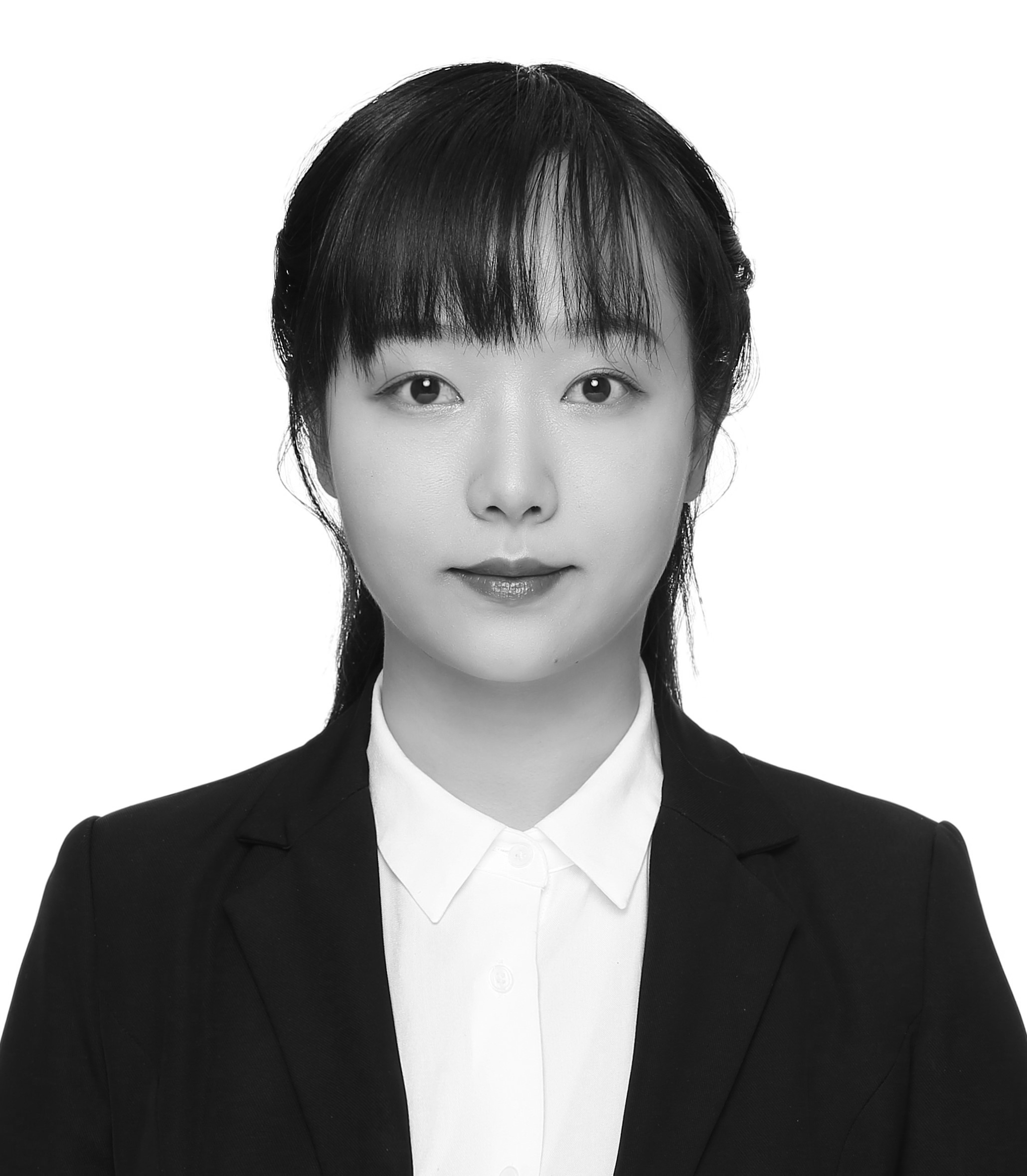}}]{Xiang~Shi} received her Bachelor's degree from the Minzu University of China, in 2014, and her PhD degree from the Institute of Computing Technology, Chinese Academy of Sciences in 2020. Currently she works in the Network Technology Laboratory at Huawei Technologies.
\end{IEEEbiography}

\begin{IEEEbiography}[{\includegraphics[width=1in,height=1.25in,clip,keepaspectratio]{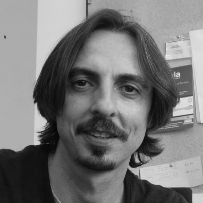}}]{Pere Barlet-Ros}
is an associate professor at Universitat Polit\`ecnica de Catalunya (UPC) and scientific director at the Barcelona Neural Networking Center (BNN-UPC). From 2013 to 2018, he was co-founder and chairman of the machine learning startup Talaia Networks. The company was acquired by Auvik Networks in 2018. He was also a visiting researcher at Endace (New Zealand), Intel Research Cambridge (UK) and Intel Labs Berkeley (USA). His research interests are in machine learning technologies for network management and optimization, traffic classification and network security. In 2014, he received the 2nd VALORTEC prize for the best business plan awarded by the Catalan Government (ACCIO) and in 2015 the Fiber Entrepreneurs award as the best entrepreneur of the Barcelona School of Informatics (FIB).
\vspace{-13cm}
\end{IEEEbiography}

\begin{IEEEbiography}[{\includegraphics[width=1in,height=1.25in,clip,keepaspectratio]{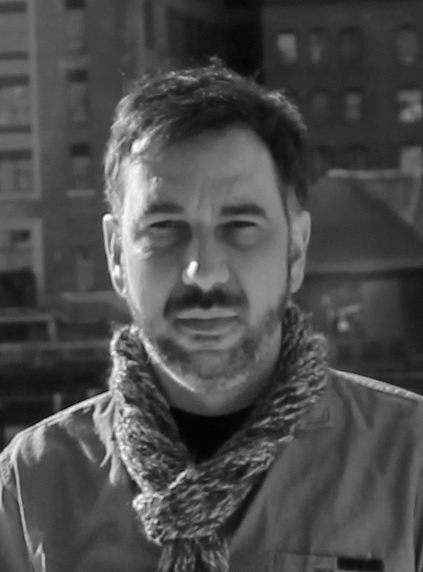}}]{Albert Cabellos-Aparicio}
is an assistant professor at Universitat Polit\`ecnica de Catalunya (UPC), where he obtained his PhD in computer science engineering in 2008. He is director of the Barcelona Neural Networking Center (BNN-UPC) and scientific director of the NaNoNetworking Center in Catalunya. He has been a visiting researcher at Cisco Systems and Agilent Technologies, and a visiting professor at the KTH, Sweden, and the MIT, USA. His research interests include the application of Machine Learning to networking and nanocommunications. His research achievements have been awarded by the Catalan Government, his university, and INTEL. He also participates regularly in standardization bodies such as the IETF.
\end{IEEEbiography}

\end{document}